\documentclass[prd,aps,nofootinbib,preprintnumbers]{revtex4}

\usepackage{graphicx,cancel,amsmath}

\begin{document}

\title{$Q$-ball dark matter and baryogenesis in high-scale inflation}

\author{Shinta Kasuya$^a$ and Masahiro Kawasaki$^{b,c}$}

\affiliation{
$^a$ Department of Mathematics and Physics,
     Kanagawa University, Kanagawa 259-1293, Japan\\
$^b$ Institute for Cosmic Ray Research, the University of Tokyo, Chiba 277-8582, Japan\\
$^c$ Kavli Institute for the Physics and Mathematics of the Universe (WPI), 
  Todai Institutes for Advanced Study, the University of Tokyo, Chiba 277-8582, Japan}

\date{August 6, 2014}

\begin{abstract}
We investigate the scenario that one flat direction creates baryon asymmetry of the universe, 
while $Q$ balls from another direction can be the dark matter in the gauge-mediated supersymmetry breaking
for high-scale inflation. Isocurvature fluctuations are suppressed by the fact that the Affleck-Dine field stays at 
around the Planck scale during inflation. We find that the dark matter $Q$ balls can be detected in 
IceCube-like experiments in the future.
\end{abstract}

\maketitle

%%%%%%%%%%%%%%%%%%%%%%%%%%%%%%%%%%%%%%
\section{Introduction}
The Affleck-Dine baryogenesis \cite{AD} is one of the promising mechanisms for creating the baryon 
asymmetry of the universe. It utilizes the scalar field which carries the baryon number. 
In the minimal supersymmetric standard model (MSSM), there are a lot of such scalar fields called flat directions
which consist of squarks (and sleptons). The scalar potential of the flat directions vanishes in the supersymmetry 
(SUSY) limit, and is lifted by SUSY breaking effects and higher order operators. 

It is well known that the $Q$ balls form during the course of the Affleck-Dine baryogenesis \cite{KuSh,EnMc,KK1}.  
In particular, $Q$ balls with large enough charge $Q$ are stable against the decay into baryons (nucleons) in the 
gauge-mediated SUSY breaking.\footnote{%%
If the charge of the $Q$ ball is small enough, the lightest SUSY particles, such as the gravitino 
\cite{KK4,KKY, gravitino} or the axino\cite{KKK}, decayed from the $Q$ ball, may be the dark matter, 
while the baryons are produced by the $Q$-ball decay as well.} %%
In this case, $Q$ balls would be dark matter of the universe, and could be 
detectable. The baryon asymmetry of the universe is explained by the baryon numbers that are not remained 
inside the $Q$ balls. It has been considered that the single flat direction transforms into dark matter $Q$ balls, 
and the baryon numbers are emitted through the $Q$-ball surface in the thermal bath of the universe
\cite{KuSh,LaSh,KK3}. Unfortunately, this simple scenario is observationally excluded \cite{KK5}.

However it is natural to consider more than one direction which take part in the Affleck-Dine mechanism. 
There are at most 29 complex degrees of freedom for $D$-flat and renormalizable $F$-flat directions which
do not involve Higgs field \cite{GKM}. Most of the directions are lifted by $n=4$ nonrenormalizable 
superpotential of the form $W=\phi^n$, which applies to the even $n$ case, where $\phi$ is the field of a flat 
direction. For the odd $n$ case, the potential is lifted by $W=H_i\phi^{n-1}$ superpotential, where 
$H_i$ is the Higgs field $H_u$ or $H_d$. In Ref.~\cite{KK5}, we considered $n=5$ and $n=6$ directions,  
where the former direction is responsible for the baryon asymmetry of the universe while the latter
forms dark matter $Q$ balls.

Recently, the BICEP2 team reported the detection of the primordial B-mode polarization of the cosmic
microwave background \cite{BICEP2}. It could be originated from gravitational waves produced 
during inflation, which leads to the large tensor-to-scalar ratio $r\simeq 0.2$ and the large Hubble parameter
during inflation $H_{\rm inf} \simeq 10^{14}$~GeV. In such high-scale inflation, in general, large baryonic 
isocurvature fluctuations will be induced in the Affleck-Dine mechanism \cite{Yokoyama, EMiso, KT,KKT}. 
In order to suppress isocurvature fluctuations sufficiently, the amplitude of the flat direction 
must be as large as the Planck scale, $M_{\rm P}=2.4\times 10^{18}$~GeV, for 
$H_{\rm inf}\simeq 10^{14}$~GeV, which can be achieved if the coupling in the nonrenormalizable 
superpotential is small enough \cite{Harigaya}.

In the present Letter, we investigate the possibility that one flat direction creates baryon asymmetry of the 
universe, while $Q$ balls from another direction can be the dark matter in the gauge-mediated SUSY breaking
for high-scale inflation.\footnote{%%
The Affleck-Dine baryogenesis and the lightest SUSY particle (LSP) dark matter production after 
high-scale inflation was considered in Ref.~\cite{Harigaya}, where all the $Q$ balls, if they form, decay 
into baryons and LSPs in both the gravity and gauge mediation.}
It is realized for two $n=4$ directions with different sizes of the couplings. The direction with larger coupling
disintegrates into the gauge-mediation type $Q$ balls, which decay into baryons before the big bang 
nucleosynthesis (BBN). The direction with smaller
coupling transforms into the new-type $Q$ balls, which are stable against the decay into baryons and become
the dark matter of the universe.\footnote{%%
The dark matter new-type $Q$ ball has a good feature that avoids astrophysical constraints by the
neutron star destruction \cite{KK5}.} %% 
These dark matter $Q$ balls could be detected by the IceCube-like experiment in the future. 

The structure of the Letter is as follows. In the next section, we review how the Affleck-Dine mechanism works
in the SUSY setup. Isocurvature fluctuations are considered in Sec.III. In Sec.IV, we show the properties
of the $Q$ balls in the gauge mediation. We estimate the baryon number and the dark matter of the universe
in Sec.V, and seek for successful scenarios to explain both the baryon asymmetry and the dark matter 
$Q$ balls in Sec.VI. Sec.VII is devoted to our conclusions.

%%%%%%%%%%%%%%%%%%%%%%%%%%%%%%%%%%%%%%
\section{Affleck-Dine mechanism}
The flat direction is a scalar field $\Phi$ which consists of squarks (and sleptons) in MSSM. 
The potential of the flat directions is flat in SUSY limit, and lifted by SUSY breaking effects and
nonrenormalizable superpotential of the form
\begin{equation}
W_{\rm NR} = \frac{\lambda \Phi^n}{n M_{\rm P}^{n-3}}, 
\label{wnr}
\end{equation}
where $\lambda$ is a coupling constant which will be determined later. In the gauge-mediated SUSY breaking,
the potential is lifted as
\begin{equation}
V = V_{\cancel{\rm SUSY}} + V_{\rm NR} = V_{\rm gauge} + V_{\rm grav} +V_{\rm NR}, 
\end{equation}
where \cite{KuSh,flat,EnMc}
\begin{eqnarray}
& & V_{\rm gauge} = \left\{ \begin{array}{ll} \frac{1}{2}m_\phi^2 \phi^2 & (\phi \ll M_{\rm m}), \\[2mm]
\displaystyle{M_F^4 \left(\log\frac{\phi^2}{2M_{\rm m}^2}\right)^2} & (\phi \gg M_{\rm m}),
\end{array}\right. \\
& & V_{\rm grav} = \frac{1}{2} m_{3/2}^2 \phi^2 \left( 1 + K\log\frac{\phi^2}{2M_*^2}\right), \\
& & V_{\rm NR} = \frac{\lambda^2 \phi^{2(n-1)}}{2^{n-1}M_{\rm P}^{2(n-3)}}. 
\end{eqnarray}
Here, $\displaystyle{\Phi=\frac{1}{\sqrt{2}}\phi e^{i\theta}}$, $m_\phi\sim O({\rm TeV})$ is a soft breaking mass, 
$M_{\rm m}$ is the messenger mass, $m_{3/2}$ is the gravitino mass, 
$K (<0)$ is a coefficient of one-loop corrections, and $M_*$ 
is the renormalization scale. $M_F$ ranges as
\begin{equation}
4 \times 10^4\, {\rm GeV} \lesssim M_F \lesssim \frac{g^{1/2}}{4\pi} \sqrt{m_{3/2}M_{\rm P}}.
\label{mflimit}
\end{equation}
Here the lower bound comes from the fact that the 125~GeV Higgs boson leads to the SUSY breaking 
parameter $\Lambda \sim 5 \times 10^5$~ GeV \cite{Hama,mGMSB}, where sparticle masses are 
estimated as $\sim g^2 \Lambda/(4\pi)^2$. 
Since the gravitino mass is much smaller than TeV scale, the potential is dominated by $V_{\rm gauge}$
for $\phi < \phi_{\rm eq}$, while $V_{\rm grav}$ overcomes $V_{\rm gauge}$ for $\phi < \phi_{\rm eq}$,
where 
\begin{equation}
\phi_{\rm eq} = \sqrt{2} \frac{M_F^2}{m_{3/2}},
\label{phieq}
\end{equation}
derived from $V_{\rm gauge}(\phi_{\rm eq})=V_{\rm grav}(\phi_{\rm eq})$. 

In the Affleck-Dine mechanism, the flat direction has large VEV during inflation. This is usually realized by
the so-called negative Hubble-induced mass term \cite{AD}
\begin{equation}
V_H = - c H^2 |\Phi|^2 = -\frac{1}{2} c H^2 \phi^2,
\end{equation}
which stems from the SUSY breaking effect by the finite inflaton energy density. Here $H$ is the Hubble 
parameter, and $c$ is a constant of $O(1)$. During inflation the flat direction stays at the minimum 
determined by the balance of $V_H$ and $V_{\rm NR}$:
\begin{equation}
\phi_{\rm min} = \sqrt{2} \left(\frac{c^{1/2} H_{\rm inf} M_{\rm P}^{n-3}}{\lambda}\right)^{1/(n-2)}.
\label{phi_min}
\end{equation}
Well after inflation when the Hubble parameter decreases as large as 
$m_{\phi,{\rm eff}} \simeq \sqrt{V''(\phi_{\rm osc})}$, the field begins oscillations. The field actually 
rotates due to the usual soft $A$ terms, and the baryon number of the universe is created.
If the field starts to oscillate at the amplitude smaller than $\phi_{\rm eq}$, the potential is dominated by 
$V_{\rm gauge}$ so that the gauge-mediation type $Q$ balls eventually form. On the other hand, 
the new-type $Q$ balls are created for $\phi_{\rm osc} > \phi_{\rm eq}$ where $V_{\rm grav}$
is larger than $V_{\rm gauge}$. It is determined by the size of $\lambda$ whether the field starts
its oscillation smaller or larger than $\phi_{\rm eq}$. The gauge-mediation type $Q$ balls form for
the flat direction with larger $\lambda$, while the new-type $Q$ balls are created for that with 
smaller $\lambda$. See Fig.1. We will consider the situation that the gauge-mediation type $Q$ balls
decay to produce the baryon asymmetry of the universe, while the new-type $Q$ balls are stable and
become the dark matter of the universe.

%%%%%%%%%%%%%%%%%%%%%%%%%%%%%%%%%%%%%%%%%
\begin{figure}[ht!]
\begin{center}
\includegraphics[width=100mm]{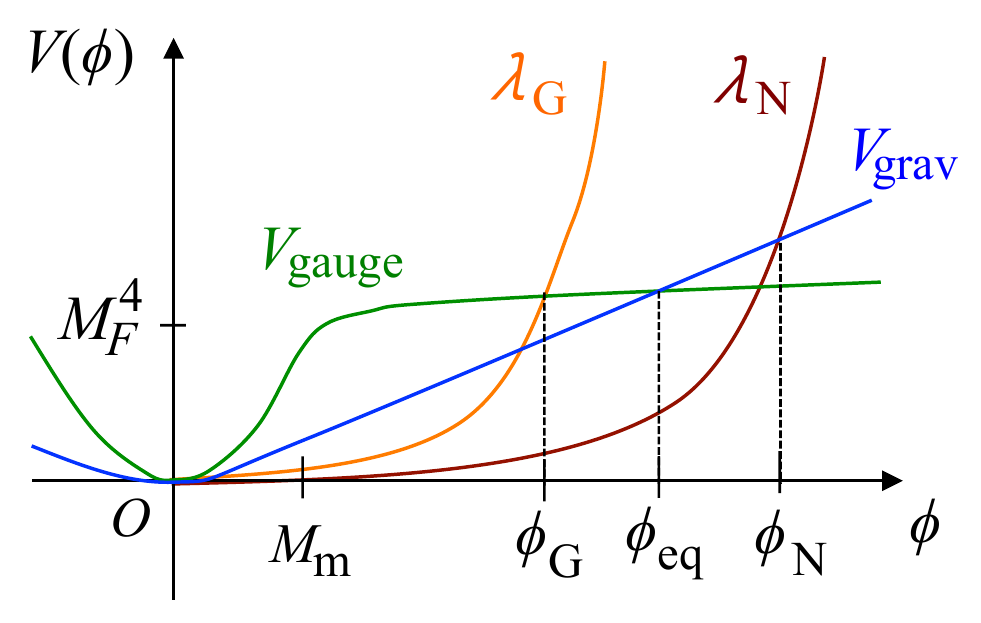} 
\caption{Potential of the flat direction. One direction is lifted by the nonrenormalizable term $V_{\rm NR}$ with 
$\lambda_{\rm G}$ (orange), while another direction is lifted by that with $\lambda_{\rm N}$ (brown).
\label{fig1}}
\end{center}
\end{figure}
%%%%%%%%%%%%%%%%%%%%%%%%%%%%%%%%%%%%%%%%%%

%%%%%%%%%%%%%%%%%%%%%%%%%%%%%%%%%%%%%%
\section{Avoiding too large isocurvature fluctuations}
During high-scale inflation the flat direction acquires quantum fluctuations along the phase direction if
there is no sizable $A$ terms, and it may lead to too large baryonic and CDM isocurvature fluctuations 
in this case. Actually, the Hubble-induced $A$ terms do not appear for most inflation models in 
supergravity \cite{KKT}. The amplitude of the baryonic isocurvature fluctuations can be estimated as
\begin{equation}
S_b = \frac{\delta\rho_b}{\rho_b} \simeq \frac{\delta n_b}{n_b} 
\simeq \frac{\delta n_\phi}{n_\phi} \simeq \delta\theta \simeq \frac{H_{\rm inf}}{2\pi\phi_{\rm inf}}.
\end{equation}
On the other hand, observational upper bound is calculated as
\begin{equation}
S_b^{\rm (obs)} \simeq \frac{\Omega_{\rm CDM}}{\Omega_b}( \beta_{\rm iso} A_s )^{1/2} 
=5.0\times 10^{-5},
\end{equation}
where $\beta_{\rm iso}=0.039$ and $\ln(10^{10} A_s)=3.098$ are respectively the primordial isocurvature 
fraction and the amplitude of the curvature power spectrum obtained in the Planck results \cite{Planck}.
It should be $S_b < S_b^{\rm (obs)}$, which results in\footnote{%
The lower limit of the amplitude of the flat direction, which is responsible for the dark matter $Q$ balls, 
is highly model-dependent. It might be at most $\Omega_{\rm CDM} / \Omega_b$ times larger than Eq.(\ref{phiinf})
for $\varepsilon=1$.}
\begin{equation}
\phi_{\rm inf} > \frac{H_{\rm inf}}{2\pi S_b^{\rm (obs)}} = 3.5 \times 10^{17} \, {\rm GeV} 
\left(\frac{H_{\rm inf}}{1.1\times 10^{14}\, {\rm GeV}}\right) \left(\frac{S_b^{\rm (obs)}}{5.0\times 10^{-5}}\right)^{-1}.
\label{phiinf}
\end{equation}
Thus the field should stay at around $M_{\rm P}$ during inflation in order to have small enough 
isocurvature fluctuations. This can be achieved for small enough $\lambda$:
\begin{equation}
\lambda \lesssim  2^{n/2-1} c^{1/2} \frac{H_{\rm inf}}{M_{\rm P}}
\underset{(n=4)}{=} 
9.3\times 10^{-5} c^{1/2} \left(\frac{H_{\rm inf}}{1.1\times 10^{14}\ {\rm GeV}}\right).
\end{equation}
Here and hereafter, $\underset{(n=4)}{=}$ means that we evaluate values for $n=4$.
Moreover, if the amplitude of the flat direction is as large as $M_{\rm P}$ during inflation, the terms such as
\begin{equation}
V_{HA} = a_H \frac{H^2 \Phi^m}{M_{\rm P}^{m-2}} +{\rm h.c.},
\end{equation}
would come from a nonminimal K\"ahler potential of the form $K = I^\dagger I \Phi^m/M_{\rm P}^m$ 
(the $F$ term of the $I$ field is responsible for inflation) 
and results in the mass scale of $O(H)$ for the phase direction \cite{Harigaya}, which further suppresses
the isocurvature fluctuations. Therefore, we adopt $\lambda \lesssim 10^{-4}$, and investigate the scenario
that the gauge-mediation type $Q$ balls from one flat direction decay into baryons, while another direction
forms new-type $Q$ balls to be dark matter of the universe in this situation.

%%%%%%%%%%%%%%%%%%%%%%%%%%%%%%%%%%%%%%
\section{$Q$ balls in the gauge mediation}
The $Q$-ball solution exists in those potential that are flatter than $\phi^2$, which is the case for SUSY.
In the gauge-mediated SUSY breaking there exist two types of the $Q$ ball. One is the gauge-mediation
type \cite{KuSh,KK3}, and the other is the new type \cite{new}. 

The former forms when the potential is dominated by $V_{\rm gauge}$
when the field starts oscillations and fragments into lumps. The charge of the formed $Q$ ball is 
estimated as \cite{KK3}
\begin{equation}
Q_{\rm G}=\beta_{\rm G} \left(\frac{\phi_{\rm osc}}{M_F}\right)^4,
\label{form-gauge}
\end{equation}
where $\phi_{\rm osc}$ is the amplitude of the field at the onset of the oscillation (rotation).
$\beta_{\rm G} \simeq 6\times 10^{-4}$ for a circular orbit ($\varepsilon=1$), while
$\beta_{\rm G} \simeq 6\times 10^{-5}$ for an oblate case ($\varepsilon\lesssim 0.1$). Here
$\varepsilon$ is the ellipticity of the field orbit. The features of this type are represented as
\begin{eqnarray}
&& M_Q \simeq \frac{4\sqrt{2}\pi}{3} \zeta M_F Q_{\rm G}^{3/4}, \\
&& R_Q \simeq \frac{1}{\sqrt{2}} \zeta^{-1} M_F^{-1} Q_{\rm G}^{1/4}, \\
\label{omegaQg}
&& \omega_Q \simeq \sqrt{2} \pi \zeta M_F Q_{\rm G}^{-1/4}, \\
&& \phi_Q \simeq \frac{1}{\sqrt{2}} \zeta M_F Q_{\rm G}^{1/4},
\end{eqnarray}
where $M_Q$ and $R_Q$ are the mass and  the size of the $Q$ ball, respectively, and $\omega_Q$ and 
$\phi_Q$ are respectively the rotation speed and the amplitude of the field inside the $Q$ ball, and 
$\zeta$ is the $O(1)$ parameter \cite{HNO, KY}. 

On the other hand, the new type of the $Q$ ball forms when the potential is dominated by $V_{\rm grav}$
at the onset of the oscillation. The charge of the new-type $Q$ ball is given by \cite{KK2, new}
\begin{equation}
Q_{\rm N} = \beta_{\rm N} \left(\frac{\phi_{\rm rot}}{m_{3/2}}\right)^2,
\label{form_new}
\end{equation}
where $\beta_{\rm N}\simeq 0.02$ \cite{Hiramatsu}. The properties of the new-type $Q$ ball are as follows: 
\begin{eqnarray}
&& M_Q \simeq m_{3/2} Q_{\rm N}, \\
&& R_Q \simeq |K|^{-1/2} m_{3/2}^{-1}, \\
\label{omegaQn}
&& \omega_Q \simeq m_{3/2}, \\
&& \phi_Q \simeq m_{3/2} Q_{\rm N}^{1/2}.
\end{eqnarray}

The charge $Q$ is in fact the $\Phi$-number,
and relates to the baryon number of the $Q$ ball as
\begin{equation}
B=bQ,
\end{equation}
where $b$ is the baryon number of $\Phi$-particle. For example, $b=1/3$ for the $udd$ direction. 

The $Q$ ball is stable against the decay into nucleons for large field amplitude when the charge
is very large. Since $\omega_Q$ corresponds to the effective mass of the field inside the $Q$ ball, 
the stability condition is given by $\omega_Q < m_{\rm D}$, where $m_{\rm D}$ is the 
mass of the decay particles. The new-type Q ball is generally stable against the decay into nucleons, 
except for the gravitino mass larger than that of nucleons.
On the other hand, the gauge-mediation type $Q$ ball is stable for $Q_{\rm G} > Q_{\rm D}$,
where 
\begin{equation}
\hspace{-5mm}
Q_{\rm D}  \equiv  4\pi^4 \zeta^4 \left(\frac{M_F}{bm_N}\right)^4 
\simeq 1.2 \times 10^{30} \left(\frac{\zeta}{2.5}\right)^4 \left(\frac{b}{1/3}\right)^{-4}
\left(\frac{M_F}{10^6 \ {\rm GeV}}\right)^4,
\label{QD}
\end{equation}
with $m_N$ being the nucleon mass. Since we need unstable $Q$ ball to produce the baryon 
number of the universe, we consider the gauge-mediation type $Q$ ball with the charge smaller than 
$Q_{\rm D}$.

The unstable gauge-mediation type $Q$ ball decays into baryons through its surface~\cite{Cohen}.\footnote{%%
The abundance of next-to-LSPs produced by the $Q$-ball decay is many orders of magnitude smaller than 
the BBN bound in our successful scenario.} %% 
The decay rate is reestimated in Refs.\cite{KY,KKY} as
\begin{equation}
\label{decayquark}
\Gamma_Q  \simeq  N_q \frac{1}{Q_{\rm G}} \frac{\omega_Q^3}{12\pi^2} 4\pi R_Q^2,
\end{equation}
where $N_q$ is the number of the decay quarks. Then the temperature at the decay is obtained as
\begin{equation}
T_{\rm D} = \left(\frac{90}{4\pi^2 N_{\rm D}}\right)^{1/4}\sqrt{\Gamma_Q^{\rm (q)} M_{\rm P}}
\simeq 16 \ {\rm MeV} \left(\frac{\zeta}{2.5}\right)^{1/2} 
\left(\frac{N_q}{18}\right)^{1/2} \left(\frac{N_{\rm D}}{10.75}\right)^{-1/4}
\left(\frac{M_F}{10^6 \ {\rm GeV}}\right)^{1/2} 
\left(\frac{Q_{\rm G}}{10^{24}}\right)^{-5/8},
\label{td}
\end{equation}
where $N_{\rm D}$ is the relativistic degrees of freedom at the decay time. Since the decay must take 
place before the BBN, the charge of the gauge-mediation type $Q$ ball has the upper limit.

%%%%%%%%%%%%%%%%%%%%%%%%%%%%%%%%%%%%%%
\section{Baryon asymmetry and the $Q$ ball dark matter}
We consider two $n=4$ flat directions with different values of $\lambda$. The direction with smaller 
$\lambda$ forms the new-type $Q$ balls to be the dark matter of the universe, while that with larger 
$\lambda$ becomes the unstable gauge-mediation type $Q$ balls to produce baryon asymmetry in 
the universe. The fields starts their oscillations when the Hubble-induced mass term $V_H$ becomes 
smaller than $V_{\cancel{\rm SUSY}}$. The amplitudes at the onset of oscillations are thus calculated as
\begin{eqnarray}
& & \phi_{\rm G,osc} 
= \sqrt{2} \lambda_{\rm G}^{-\frac{1}{n-1}} \left(M_F^2 M_{\rm P}^{n-3} \right)^{\frac{1}{n-1}}
\underset{(n=4)}{=}  \sqrt{2} \lambda_{\rm G}^{-\frac{1}{3}} \left(M_F^2 M_{\rm P} \right)^{\frac{1}{3}},
\label{phig}
\\
& & \phi_{\rm N,osc}
= \sqrt{2}   \lambda_{\rm N}^{-\frac{1}{n-2}} \left(m_{3/2} M_{\rm P}^{n-3} \right)^{\frac{1}{n-2}}
\underset{(n=4)}{=}  \sqrt{2}   \lambda_{\rm N}^{-\frac{1}{2}} \left(m_{3/2} M_{\rm P} \right)^{\frac{1}{2}},
\label{phin}
\end{eqnarray}
for the gauge-mediation type and the new type, respectively.  
Therefore, the charges of both types of the $Q$ balls are obtained as
\begin{eqnarray}
& & Q_{\rm G} = \beta_{\rm G} \left(\frac{\phi_{\rm G,osc}}{M_F}\right)^4 
= \frac{4\beta_{\rm G}}{\lambda_{\rm G}^{\frac{4}{n-1}}} 
\left(\frac{M_{\rm P}}{M_F}\right)^{\frac{4(n-3)}{n-1}}
\underset{(n=4)}{=} \frac{4\beta_{\rm G}}{\lambda_{\rm G}^{\frac{4}{3}}} 
\left(\frac{M_{\rm P}}{M_F}\right)^{\frac{4}{3}} \quad ({\rm gauge\mathchar`- mediation \, type}),
\\
& & Q_{\rm N}=\beta_{\rm N} \left(\frac{\phi_{\rm N,osc}}{m_{3/2}}\right)^2
= \frac{2\beta_{\rm N}}{\lambda_{\rm N}^{\frac{2}{n-2}}} \left(\frac{M_{\rm P}}{m_{3/2}}\right)^{\frac{2(n-3)}{n-2}} 
\underset{(n=4)}{=} \frac{2\beta_{\rm N}}{\lambda_{\rm N}}\left(\frac{M_{\rm P}}{m_{3/2}}\right)
\quad ({\rm new \, type}).
\label{QN}
\end{eqnarray}

Since all the charges (the baryon numbers) are released by the decay of the gauge-mediation type $Q$ balls,
the baryon number is given by
\begin{equation}
Y_b = \frac{9}{8\sqrt{2}} b \frac{T_{\rm RH}\phi_{\rm G,osc}^3}{M_F^2M_{\rm P}^2}
= \frac{9}{4} b \lambda_{\rm G}^{-\frac{3}{n-1}} T_{\rm RH} M_F^{-\frac{2(n-4)}{n-1}} M_{\rm P}^\frac{n-7}{n-1}
\underset{(n=4)}{=} \frac{9}{4} b \lambda_{\rm G}^{-1} T_{\rm RH} M_{\rm P}^{-1},
\end{equation}
where $T_{\rm RH}$ is the reheating temperature after inflation. In this $n=4$ case,
$T_{\rm RH}$ is independent of $M_F$ and solely determined by $\lambda_{\rm G}$ as
\begin{equation}
T_{\rm RH} = \frac{4}{9} b^{-1} \lambda_{\rm G} Y_b M_{\rm P} = 3.2 \times 10^4 \ {\rm GeV} 
\left(\frac{b}{1/3}\right)^{-1} \left(\frac{\lambda_{\rm G}}{10^{-4}}\right)
\left(\frac{Y_b}{10^{-10}}\right).
\label{Trh}
\end{equation}

At the same time, the dark matter is composed by stable new-type $Q$ balls. Their abundance is
estimated as
\begin{equation}
\frac{\rho_Q}{s} = \frac{9}{8} T_{\rm RH} \left(\frac{\phi_{\rm N,osc}}{M_{\rm P}}\right)^2 
= \frac{9}{4} \lambda_{\rm N}^{-\frac{2}{n-2}}T_{\rm RH} \left(\frac{m_{3/2}}{M_{\rm P}}\right)^\frac{2}{n-2}
\underset{(n=4)}{=} \frac{9}{4} \lambda_{\rm N}^{-1}T_{\rm RH} \left(\frac{m_{3/2}}{M_{\rm P}}\right),
\label{rhoQth}
\end{equation}
which is rewritten as
\begin{equation}
m_{3/2} = \left(\frac{2}{3}\right)^{n-2} \lambda_{\rm N} \left(\frac{\rho_Q}{s}\right)^\frac{n-2}{2}
T_{\rm RH}^{-\frac{n-2}{2}} M_{\rm P}
\underset{(n=4)}{=} \left(\frac{2}{3}\right)^2 \lambda_{\rm N} \left(\frac{\rho_Q}{s}\right)
\frac{M_{\rm P}}{T_{\rm RH}}.
\label{m32BDM}
\end{equation}
We must have
\begin{equation}
\frac{\rho_Q}{s} = 5.4 m_N Y_b = 5.4 \times 10^{-10} \ {\rm GeV} \left(\frac{Y_b}{10^{-10}}\right),
\label{rhoQobs}
\end{equation}
in order to have right amount of the dark matter. For explaining both the baryon asymmetry and 
the dark matter of the universe, we have
\begin{equation}
m_{3/2} = 1.8 \ {\rm MeV} 
\left(\frac{b}{1/3}\right)
\left(\frac{\lambda_{\rm N}}{10^{-7}}\right) 
\left(\frac{\lambda_{\rm G}}{10^{-4}}\right)^{-1},
\label{BDM}
\end{equation}
where we insert Eqs.(\ref{Trh}) and (\ref{rhoQobs}) with $n=4$ into Eq.(\ref{m32BDM}). 
We call this the B-DM relation in the following.

%%%%%%%%%%%%%%%%%%%%%%%%%%%%%%%%%%%%%%%%%%%
\section{Constraints on model parameters and detection of dark matter $Q$ balls }
Let us now investigate how the B-DM relation (\ref{BDM}) is satisfied. We consider two flat 
directions with different values of $\lambda$. Since isocurvature fluctuations should be 
suppressed, it is necessary to have $\lambda_{\rm N} < \lambda_{\rm G} \lesssim 10^{-4}$. 
There are several conditions to be hold for successful scenario. (a) The flat direction with $\lambda_{\rm G}$ 
disintegrates into the gauge-mediation type $Q$ balls, so that $\phi_{\rm G,osc} < \phi_{\rm eq}$, 
while (b) the new-type $Q$ balls form from the direction with $\lambda_{\rm N}$ for 
$\phi_{\rm N,osc} > \phi_{\rm eq}$. From Eqs.(\ref{phieq}) and (\ref{phig}), the condition (a) results in
\begin{equation}
m_{3/2} < \lambda_G^{\frac{1}{n-1}} \left(\frac{M_F^{2(n-2)}}{M_P^{n-3}}\right)^{\frac{1}{n-1}}
\underset{(n=4)}{=} 3.47 \, {\rm GeV} \left(\frac{M_F}{10^6\, {\rm GeV}}\right)^{4/3} 
\left(\frac{\lambda_{\rm G}}{10^{-4}}\right)^{1/3},
\label{phigosc}
\end{equation}
while the condition (b) leads to
\begin{equation}
m_{3/2} > \lambda_N^{\frac{1}{n-1}} \left(\frac{M_F^{2(n-2)}}{M_P^{n-3}}\right)^{\frac{1}{n-1}}
\underset{(n=4)}{=} 0.347 \, {\rm GeV} \left(\frac{M_F}{10^6\, {\rm GeV}}\right)^{4/3} 
\left(\frac{\lambda_{\rm N}}{10^{-7}}\right)^{1/3}.
\label{phinosc}
\end{equation}

The gauge-mediation type $Q$ ball (c) should be unstable and (d) decays before the BBN.
The former condition is expressed, from Eqs.(\ref{form-gauge}) and (\ref{omegaQg}), as
\begin{equation}
M_F > \left(\frac{b\beta_{\rm G}^{1/4}}{\pi \zeta}\right)^\frac{n-1}{2(n-2)} \lambda_G^{-\frac{1}{2(n-2)}}
\left(m_N^{n-1} M_{\rm P}^{n-3}\right)^\frac{1}{2(n-2)}
\underset{(n=4)}{=} 9.16 \times 10^3 \ {\rm GeV} \left(\frac{\lambda_{\rm G}}{10^{-4}}\right)^{-1/4}
\left(\frac{b}{1/3}\right)^{3/4}, 
\label{cond-unst}
\end{equation}
with $\beta_{\rm G} = 6\times 10^{-4}$ and $\zeta=2.5$ being used. 
The latter condition comes from $T_{\rm D} > 1$~MeV, which leads to
\begin{equation}
M_F > 7.19 \times 10^2 \ {\rm GeV} \left(\frac{\lambda_G}{10^{-4}}\right)^{-5/8},
\label{cond-bbn}
\end{equation}
for $n=4$, from Eqs.(\ref{td}) and (\ref{form-gauge}). Since Eqs.(\ref{cond-unst}) and (\ref{cond-bbn}) 
have different dependences on $\lambda_{\rm G}$, the latter gives more stringent constraint for
$\lambda_{\rm G} < 1.13 \times 10^{-7} \left(\frac{b}{1/3}\right)^{-2}$.

\begin{figure}[ht!]
\begin{center}
\includegraphics[width=150mm]{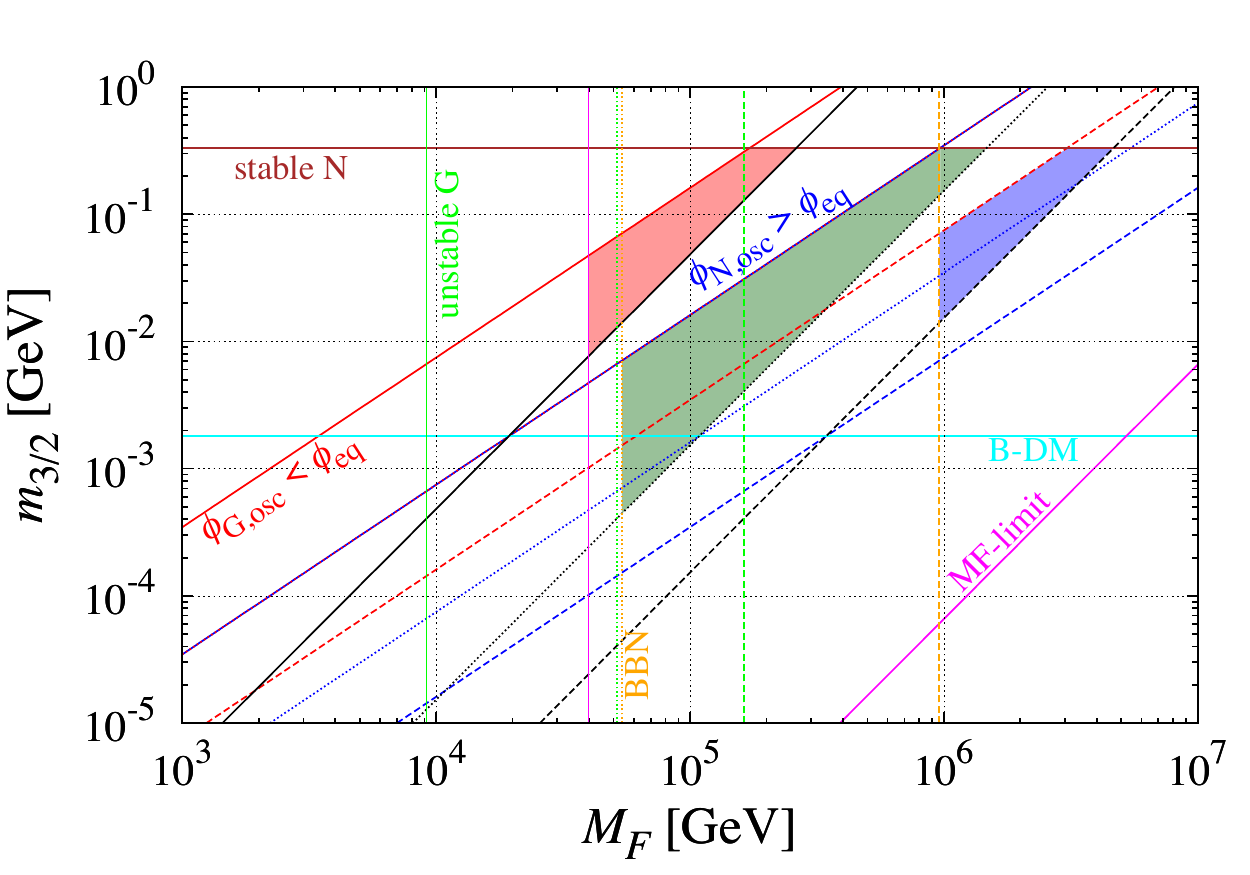} 
\caption{B-DM relation (\ref{BDM}) is displayed in a cyan line for  $\lambda_{\rm G}/\lambda_{\rm N}=10^3$. 
There are the conditions (a) Eq.(\ref{phigosc}) in red lines, (b) Eq.(\ref{phinosc}) in blue lines, 
(c) Eq.(\ref{cond-unst}) in green lines, (d) Eq.(\ref{cond-bbn}) in orange lines, 
and (e) Eq.(\ref{cond-st}) in a brown line. Also shown is the constraint (\ref{mflimit}) in magenta lines. 
We obtain red, green, and blue hatched regions respectively for $\lambda_{\rm G} = 10^{-4}$, $10^{-7}$, 
and $10^{-9}$ with varying $\lambda_{\rm N}$. 
\label{fig2}}
\end{center}
\end{figure}

In addition, (e) the new-type $Q$ ball must be stable to be the dark matter of the universe. From 
Eq.(\ref{omegaQn}), $\omega_Q  < m_D =b m_N$ is nothing but the upper bound on the gravitino mass:
\begin{equation}
m_{3/2} < 0.333 \ {\rm GeV} \left(\frac{b}{1/3}\right).
\label{cond-st}
\end{equation}
There is also the condition (\ref{mflimit}). The upper bound is not restrictive, but the lower bound limits
the smaller $m_{3/2}$ for $\lambda_{\rm G} \lesssim 10^{-7}$, as seen in Fig.~\ref{fig2}.

In Fig.~\ref{fig2}, we display the B-DM relation (\ref{BDM}) in the cyan line with 
$\lambda_{\rm G}/\lambda_{\rm N}=10^3$.
Also shown are the conditions (a) in red lines, (b) in blue lines, (c) in green lines, (d) in orange lines, 
and (e) in a brown line. Solid, dotted, and dashed lines correspond for $\lambda_{\rm G}=10^{-4}$, $10^{-7}$,  
and  $10^{-9}$, respectively. Varying the value of $\lambda_{\rm N}$, we obtain the allowed range for B-DM 
relation in red, green, and blue hatched regions for $\lambda_{\rm G}=10^{-4}$, $10^{-7}$, and $10^{-9}$, 
respectively. Notice that the condition (\ref{phinosc}) depends on $\lambda_{\rm N}$ so that the allowed 
range of $M_F$ for the B-DM relation (\ref{BDM})  becomes narrower as
$\lambda_{\rm N}$ grows larger. This is shown by black lines in the figure.
The allowed region becomes smaller as $\lambda_{\rm G}$ gets smaller, and 
disappears for $\lambda_{\rm G} < 1.46 \times 10^{-11}$. 
One can see that the allowed region appears only for 0.404~MeV $< m_{3/2}<$ 0.33~GeV,
and typically for $4\times 10^4 \, {\rm GeV} \lesssim M_F \lesssim 10^7$~GeV.

%%%%%%%%%%%%%%%%%%%%%%%%%%%%%%%%%%%%%%%%%
%%%%%%%%%%%%%%%%%%%%%%%%%%%%%%%%%%%%%%%%%%

\begin{figure}[ht!]
\begin{center}
\includegraphics[width=150mm]{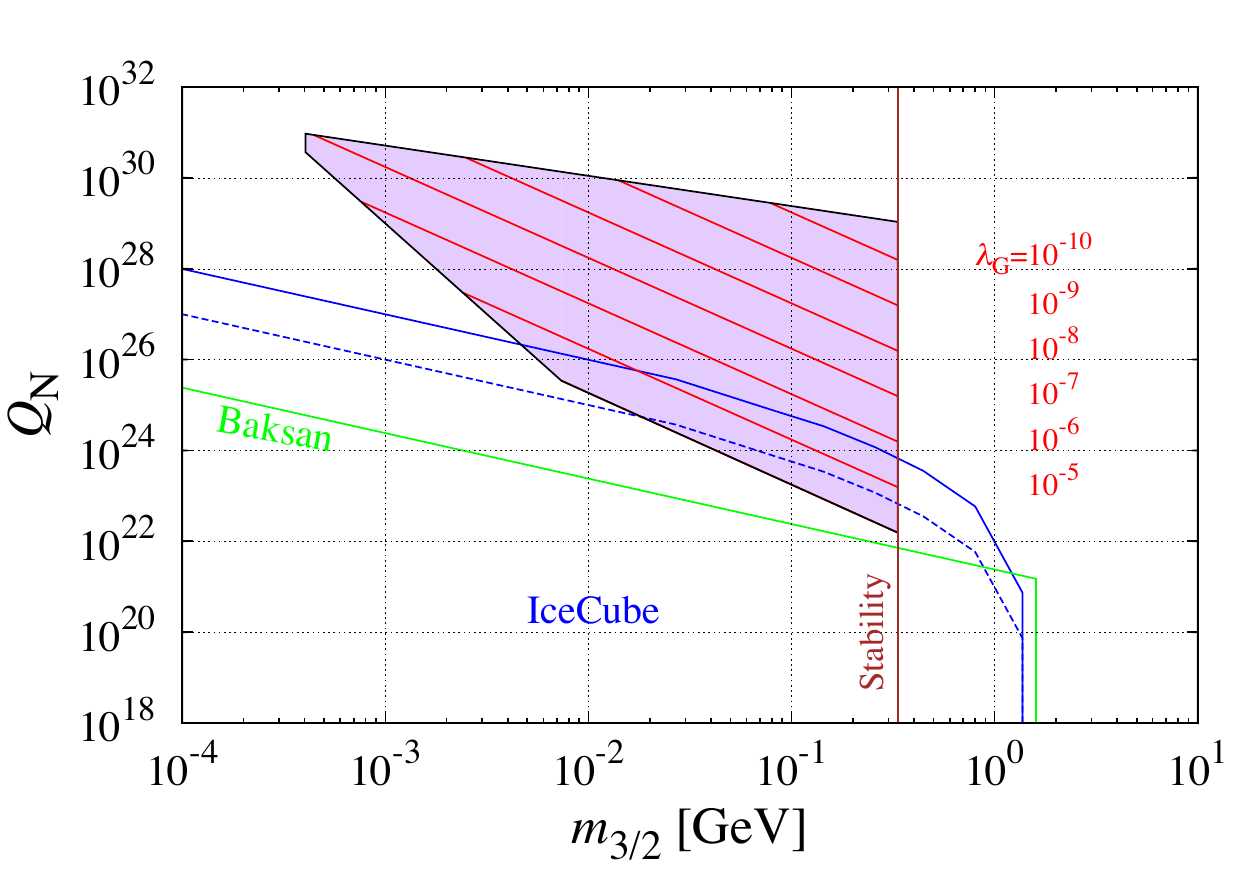} 
\caption{Theoretically expected region for the new-type $Q$ ball as dark matter. 
Iso-$\lambda_{\rm G}$ contours are shown in red lines. Observationally excluded region by the
Baksan experiment is shown in green lines, while that by the IceCube is shown in blue lines.
(Updated from the published version: Flux limit is extended to larger cross section for the IceCube data.)
\label{fig3}}
\end{center}
\end{figure}

Now we estimate how large the new-type $Q$ ball will be for our successful scenario. 
The charge of the new-type $Q$ ball is calculated as
\begin{equation}
Q_N \simeq 1.06\times 10^{28}  \left(\frac{b}{1/3}\right) 
\left(\frac{\lambda_{\rm G}}{10^{-4}}\right)^{-1} 
\left(\frac{m_{3/2}}{0.404\, {\rm MeV}}\right)^{-2},
\end{equation}
from Eqs.(\ref{QN}) and (\ref{BDM}).
We plot the expected region of the charge of the dark matter $Q$ ball for 
$10^{-4} \lesssim \lambda_{\rm G} \lesssim 10^{-11}$ in Fig.~\ref{fig3}. We display the observationally 
excluded region by Baksan in green lines (lower-left region is excluded) \cite{Arafune, new}.
The recent IceCube experiment reported the limits on subrelativistic magnetic monopoles \cite{IC}, 
which can be applied to the $Q$-ball search. Since we do not know the detection efficiency $\eta$, 
we show the constraints for $\eta=1$ (0.1) in blue solid (dashed) lines. We can see that the observation
has just stepped into the theoretically expected region, and we can imagine that the dark matter $Q$ balls
could be detected directly in the future, which will be a smoking gun for the Affleck-Dine mechanism.

%%%%%%%%%%%%%%%%%%%%%%%%%%%%%%%%%%%%%%%%%
%%%%%%%%%%%%%%%%%%%%%%%%%%%%%%%%%%%%%%%%%%

%%%%%%%%%%%%%%%%%%%%%%%%%%%%%%%%%%%%%%
\section{Conclusion}
We have investigated the variant of the Affleck-Dine mechanism in high-scale inflation. The scenario is that
one flat direction is responsible for the baryon asymmetry of the universe, while another explains the dark 
matter in the gauge-mediated SUSY breaking in high-scale inflation. The former disintegrates into the 
gauge-mediation type $Q$ balls, which are unstable and decay into baryons. The latter forms the 
new-type $Q$ balls to become dark matter of the universe. 

Large isocurvature fluctuations are usually produced in high-scale inflation. In order to suppress them, the flat
directions must have amplitudes as large as $M_{\rm P}$ during inflation. This is achieved by small coupling
constants $\lambda$ in nonrenormalizable superpotential (\ref{wnr}) such that $\lambda \lesssim 10^{-4}$.
We have found that the scenario actually works for two $n=4$ directions with different sizes of $\lambda$. 
It is necessary for successful scenario to have $10^{-11} \lesssim \lambda_{\rm G} \lesssim 10^{-4}$. In this case, 
$4\times 10^4\, {\rm GeV} \lesssim M_F \lesssim 10^7\, {\rm GeV}$, MeV $\lesssim m_{3/2} \lesssim$ GeV, and
$10^{23} \lesssim Q_{\rm N} \lesssim 10^{31}$. Smallest $Q_{\rm N}$ region may already be excluded,
but the dark matter $Q$ ball with $Q\gtrsim 10^{24}$ could be detected in the future by the IceCube-like detector.

%%%%%%%%%%%%%%%%%%%%%%%%%%%%%%%%%%%%%%%%%%
\section*{Acknowledgments}
We thank Masahiro Ibe and Tsutomu Yanagida for useful discussion. 
The work is supported by Grant-in-Aid for Scientific Research  
23740206 (S.K.) and 25400248 (M.K.) 
from the Ministry of Education, Culture, Sports, Science, and 
Technology in Japan, and also by World Premier International Research 
Center Initiative (WPI Initiative), MEXT, Japan.

%%%%%%%%%%%%%%%%%%%%%%%%%%%%%%%%%%%%

%%%%%%%%%%%%%%%%%%%%%%%%%%%%%%%%%%%%

\end{document}